\documentstyle[preprint,aps]{revtex}
\begin{document}
\input epsf
\preprint{UMHEP-432}
\title{
Factorization in non-leptonic 
decays of heavy mesons.}
\author{John F. Donoghue, Alexey A. Petrov}
\address{Department of Physics and Astronomy \\
         University of Massachusetts \\
         Amherst, MA 01003 }
\maketitle
\begin{abstract}
We propose a QCD-based model for calculation of the
non-perturbative corrections to the factorization
approximation in the decays of heavy mesons.
In the framework of the model, factorization in pseudoscalar
transitions holds 
exactly at the leading order leaving the opportunity to
calculate non-leading corrections consistently. 
\end{abstract}

\thispagestyle{empty}
\newpage
\setcounter{page}{1}

\section{Introduction}
The factorization approximation is widely used in the heavy quark
physics. As a calculational scheme, it is very 
convenient and there are numerous models based on it. However, the
theoretical basis for the approximation itself is not completely 
understood. The usual motivation for it is a 
color transparency argument due to Bjorken
\cite{bj} which suggests that 
the quark pair produced in the decay does not have enough time to
evolve to the real-size hadronic entity, being a small size
bound state with a small chromomagnetic moment. That is taken to imply
that
the QCD interaction is suppressed. For the heavy-to-heavy-light 
transition (we shall hereafter refer to the color allowed
decay $B \to D^+ \pi^-$ for the sake of clarity)
the matrix elements in question are 
\begin{eqnarray} \label{firfor}
& \left.
{\cal M} (B(p_B) \to D(p_D) \pi(q))=
\langle \pi D | {\cal H}(0) | B \rangle =
\frac{G_F}{\sqrt{2}} V_{cb} V_{du} \Bigl \{ 
\Bigl( C_1 + \frac{1}{N_c} C_2 \Bigr ) {\cal M_{\text{1}}} +
2 C_2 {\cal M_{\text{8}}}  \Bigr \} =
\right. &
\nonumber \\ 
& \left.
\frac{G_F}{\sqrt{2}} V_{cb} V_{du} \Bigl \{
\Bigl( C_1 + \frac{1}{N_c} C_2 \Bigr ) 
\langle \pi D | \bar c (0) \gamma_\mu ( 1 + \gamma_5)
b (0) ~ \bar d (0) \gamma^\mu ( 1 + \gamma_5) u (0) | B \rangle
+ \right.
\nonumber \\
& \left. 
2 C_2 \langle \pi D | \bar c (0) \gamma_\mu ( 1 + \gamma_5) t^a
b (0) ~ \bar d (0) \gamma^\mu ( 1 + \gamma_5) t_a u (0) | 
B \rangle \Bigr \} \right. &
\end{eqnarray}
Here the usual notations are used
\begin{eqnarray}
& \left.
{\cal H} =  C_1 O_1 + C_2 O_2 ~, ~~
\langle \pi D |~ O_1 ~ | B \rangle = {\cal M_{\mbox{1}}}, ~~~~
\langle \pi D |~ O_2 ~ | B \rangle = {1 \over N_c}{\cal M_{\mbox{1}}}
+ 2{\cal M_{\mbox{8}}} \right. &
\nonumber \\
& \left.
O_1 = \bar c_i \gamma_\mu ( 1 + \gamma_5) b^i~ 
\bar d_k \gamma^\mu ( 1 + \gamma_5) u^k ~~ \mbox{and} ~~~
O_2 = \bar c_i \gamma_\mu ( 1 + \gamma_5) b^k ~
\bar d_k \gamma^\mu ( 1 + \gamma_5) u^i 
\right. &
\end{eqnarray}
where we have explicitly written out color indices.
In the usual factorization approximation one simply separates 
out currents by inserting the vacuum state and disregards any QCD 
interactions between them on the basis of Bjorken's conjecture.
That implies that the  ${\cal M_{\text{8}}}$ 
contribution in (\ref{firfor}) is dropped, as it has no factorizable
color singlet form.
New non-factorizable corrections appearing in the
 ${\cal M_{\text{8}}}$ matrix element
can arise if a hard or soft gluon is emitted or
absorbed by the escaping pair. 
Obviously, a more quantitative estimate of these contributions is
needed.

The first attempt to put a factorization approximation on the 
solid theoretical base was made in \cite{dg}. 
The authors considered a limit of large energy transferred to 
light quarks 
\begin{equation} \label{energy}
E = v \cdot q = \frac{m_b^2 - m_c^2}{2m_b} 
\sim {m_b}, 
\end{equation}
as ${m_b \to \infty}$. Here $v$ is the velocity of the decaying heavy quark
$p_b = m_b v + k$ and $q$ is the momentum transferred to the 
light quark pair. In this
limit the energy scales as a heavy quark mass. 
Proposing a Large Energy Effective Theory (LEET) 
as an extension of the Heavy Quark Effective Theory and making use of
a convenient gauge condition $n \cdot A = 0$ (with $n$ being a 
null vector, $n^2$=0) they proved factorization for the physical 
amplitude dominated by collinear quarks. It is not clear, however,
how the pair of collinear quarks hadronizes into the pion.
Also, the validity of the LEET itself is not yet established.
  
The important step towards the sought theoretical description 
was made in \cite{bs}. The authors considered 
a specific small velocity limit where
$m_b - m_c \sim const$ which implies different scaling of 
(\ref{energy}): $E \sim const$ as ${m_b \to \infty}$. 
It turns out that this limit is
theoretically ``clean'' for the application of the QCD sum rule
method and provides a theoretical justification of the
``rule of discarding $1/N_c$ corrections'' \cite{buras} on
dynamical grounds. Unfortunately, the actual $B$ decays are at the 
borderline of the formalism, bringing large uncertainties into the
estimation of the matrix elements. 

In what follows we develop a QCD-based model which works in the 
limit $E \sim {m_b}$ as ${m_b \to \infty}$. The idea is to study the
propagation of the light quarks emerging to form the final hadron
while passing through the (perturbative or
non-perturbative) gluonic fields present in the decay. Interestingly,
results can be obtained which are independent of the nature of the
gluonic fields. 
The light quark pair, produced at $x=0$,  
interacts strongly with the background 
gluonic field 
while escaping from the point they were produced and hadronizes 
into the pion. Since the energy transferred to
the quark pair is large it is not necessary  that the hadronization
occurs at $x \sim 0$. 
As a result, it is not sufficient to take the operators
of the lowest dimension in the Operator Product Expansion (OPE)
but rather one must sum up a whole series of operators. 
The novel technical point of our method is to use introduce 
``generating (distribution) functions'' which incorporate this
infinite series of matrix elements, and then to use 
Heavy Quark Symmetry to restrict
the form of the matrix elements enough that useful statements can be
made. In principle, these functions could be modeled or perhaps fixed
by other measurements. However,  at the leading order in 
$q^2=0, x^2 \sim 0$ the color octet matrix element vanishes due to 
cancellations and factorization holds. The method should be able to be
extended beyond leading order, allowing one to also see corrections to
factorization.  

\section{The Octet Contribution}

The amplitude for the singlet or octet part is 

\begin{eqnarray} \label{master}
{\cal M}_{1,8} & = &
\langle D \pi | \bar c (0) \gamma_\mu ( 1 + \gamma_5) T b(0)
\bar d \gamma^\mu ( 1 + \gamma_5) T u 
| B \rangle     \nonumber \\
 & = &
 \langle D | \bar c (0) \gamma_\mu ( 1 + \gamma_5) T b (0)
{\cal P}_\mu
 | B \rangle
\end{eqnarray}
with $T=\{ 1,t^a \}$ being the appropriate color matrix. 
The quantity ${\cal P}_\mu$ describes the transition of the light
quarks into a pion in the presence of the gluonic fields which are
either internal to the B,D hadrons or produced during the transition. 
The idea is to treat these gluonic fields as external fields 
with respect to the light quark pair produced in the decay of virtual 
$W$-meson. From the diagrams Fig.1 one may write down the
expression for one gluon matrix element of the produced quark pair
in this general background field. This gives a contribution to the 
octet matrix element only:

\begin{eqnarray} \label{onegluon}
& \left.
{\cal P}_{\mu} = -(i g_s) \int d^4 x \langle \pi | 
\bar d (x) \rlap /{{\cal A^{\text{b}}}} (x) t^b_{ik}
S_F(x,0) \gamma_\mu (1 + \gamma_5) t^a_{ki}
u (0) | 0 \rangle - 
\right. &
\nonumber \\ 
& \left.
(i g_s) \int d^4 x \langle \pi | 
\bar d (0) \gamma_\mu (1 + \gamma_5) t^a_{ki}
S_F(0,x) \rlap /{{\cal A^{\text{b}}}} (x) t^b_{ik}
u (x) | 0 \rangle  
\right. &
\end{eqnarray}
where we set $ {\cal A}_\mu \equiv A_\mu$ = the external 
field. As one can
see, in this approximation we neglect higher Fock state 
contributions.
Our normalization is such that $Tr~t^a t^b = \delta^{ab}/2$.
Thus we can rewrite the equation (\ref{onegluon}) as
\begin{eqnarray} \label{onegl}
& \left.
{\cal P}_{\mu}= - \frac{i g_s}{2} \int d^4 x \Bigl ( 
Tr \rlap /{A^a} (x) S_F(x,0) \gamma_\mu (1 + \gamma_5)
G(0,x) +  \right. &
\nonumber \\
& \left.
Tr \gamma_\mu (1 + \gamma_5) S_F(0,x) \rlap /{A^a} (x)
G(x,0) \Bigr )
\right. &
\end{eqnarray} 
with non-local correlators
\begin{eqnarray}
& \left.
G(0,x) =  \langle \pi | T u (0) \bar d (x) | 0 \rangle 
\right. &
\nonumber \\
& \left.
G(x,0) =  \langle \pi | T u (x) \bar d (0) | 0 \rangle
\right. &
\end{eqnarray}
Shifting the $x$-dependence of the second correlator
\begin{equation}
\langle \pi | T u (x) \bar d (0) | 0 \rangle =
\langle \pi | T e^{iPx} u (0) e^{-iPx} 
\bar d (0) e^{iPx} e^{-iPx} | 0 \rangle = e^{iqx} G(0,-x) \ \ .
\end{equation}
We shall work in the 
Fock-Shwinger gauge \cite{fs} which gives us the advantage of
expressing
the $A_\mu$ in terms of gauge-invariant quantities
(such as the gluon stress tensor $G_{\rho \mu}^a$):
\begin{equation} \label{amu}
x_\mu \cdot A^\mu = 0, ~~ 
A_\mu^a = \int_0^1 d \alpha ~ \alpha
~ x_\rho G_{\rho \mu}^a (\alpha x) \ \ .
\end{equation} 
We treat the light quarks as massless, so the bare propagator of a light 
quark is given by
\begin{equation} \label{bare}
S_F(x-y) = \frac{1}{2 \pi^2} \frac{\rlap /{x} - \rlap /{y}}{((x-y)^2)^2}
\end{equation}
This allows us to rewrite (\ref{onegl}) as

\begin{eqnarray}
& \left. -i {\cal P}_{\mu}= 
-i {\cal P}_{1~\mu} - i  {\cal P}_{2~\mu} =
\frac{(-i g_s)}{4 \pi^2} \int d^4 x 
\int_0^1 \alpha d \alpha \Bigl[ 
\frac{x_\rho x_t}{x^4} G_{\rho s}^a (\alpha x)   
Tr \sigma_{st} \gamma_\mu (1 + \gamma_5)
G(0,x) -  \right. & \nonumber \\
& \left.
\frac{x_u x_\rho}{x^4} G_{\rho n}^a (\alpha x)
Tr \gamma_\mu (1 + \gamma_5) \sigma_{un} G(0,-x) e^{iqx} \Bigr]
\right. &
\end{eqnarray} 
The correlator $G(0,\pm x)$ entering the expression above is 
non-perturbative and nonlocal but can be  
related to the wave function of the forming pion.
Following \cite{brodsky} we expand it near the light cone ($x^2 \sim 0$).
The general form of expansion is

\begin{eqnarray} \label{expansion}
T u (0) \bar d (x) =
\sum_n C_n (x^2- i \epsilon) \Gamma_\alpha
x^{\mu_1} ... x^{\mu_n} ( \bar d~ \Gamma^\alpha 
D_{\mu_1}...D_{\mu_n} u ) + \nonumber \\
\sum_{n,m} C_{nm} (x^2 - i \epsilon) \Gamma_\alpha
x^{\mu_1} ... x^{\mu_n} ( \bar d~ \Gamma^\alpha 
D_{\mu_1}...D_{\mu_m} A_{\mu_{m+1}}...A_{\mu_n} u )
\end{eqnarray}
with $\Gamma_\alpha$ being the full set of Dirac matrices. We would 
like to note, however, that since $G(0,x)$ (or $G(x,0)$)
contributes to the trace, only $\gamma_\alpha$ and $\gamma_5 \gamma_\alpha$ 
terms enter the final expression for $G(0,x)~(G(x,0))$.
In Eq. (\ref{expansion}), the non-gauge invariant terms are exactly zero
by the FS gauge condition (\ref{amu}) which allows us to use analysis similar to 
\cite{brodsky}. Moreover, in this gauge $x^\mu D_{\mu_i} \equiv 
x^\mu \partial_{\mu_i}$ in
(\ref{expansion}). Thus,

\begin{eqnarray}
\langle \pi | T u (x) \bar d (0) | 0 \rangle =
\sum_n C_n \Gamma_\alpha x^{\mu_1} ... x^{\mu_n}
\langle \pi | \bar d \Gamma^\alpha 
D_{\mu_1}...D_{\mu_n} u  | 0 \rangle
\end{eqnarray}
Parameterizing unknown matrix elements in the conventional form
\begin{equation}
\langle \pi | \bar d ~ \Gamma^\alpha 
D_{\mu_1}...D_{\mu_n} u  | 0 \rangle =
B_n q^\alpha q_{\mu_1}...q_{\mu_n}
\end{equation}
where $q$ is a momentum of the pion we arrive at
the simple form for the $G$-correlators 
\begin{eqnarray} \label{gcorr}
\langle \pi | T u (0) \bar d (x) | 0 \rangle =
(\Gamma_\alpha q^\alpha ) \int dz e^{izxq} \phi(z) \nonumber \\
\langle \pi | T u (0) \bar d (-x) | 0 \rangle =
(\Gamma_\alpha q^\alpha ) \int dz e^{i(1-z)xq} \phi(z) 
\end{eqnarray}
with the function $\phi(z)$ introduced through its moments
\begin{equation}
\int dz z^n \phi(z) = (-i)^n n! C_n B_n
\end{equation}
The zeroth moment 
gives the overall normalization of $\phi(z)$:
$\int dz ~ \phi(z) = f_\pi / (2 \sqrt{N_c}) $ with
$f_\pi$ being the decay constant of $\pi$-meson. 
Inserting (\ref{gcorr}) into the (\ref{onegl}) we obtain
\begin{eqnarray}
-i {\cal P}_{1\mu} = \frac{g_s}{\pi^2} 
\int dz d^4 x
\frac{\phi(z)}{x^4} e^{izxq}
\Bigl \{ x_\mu q \cdot A^b(x) -
A^b_\mu (x) (x \cdot q) -
i \epsilon_{st \mu \alpha} A^b_s(x) x_t q_\alpha \Bigr \}
\nonumber \\
-i {\cal P}_{2 \mu} = - \frac{g_s}{\pi^2} 
\int dz d^4 x
\frac{\phi(z)}{x^4} e^{i(1-z)xq}
\Bigl \{ x_\mu q \cdot A^b(x) -
A^b_\mu (x) (x \cdot q) +
i \epsilon_{st \mu \alpha} A^b_s(x) x_t q_\alpha \Bigr \}
\end{eqnarray}
This gives for the octet amplitude of $B \to D \pi$ transition:
\begin{eqnarray} \label{octet}
{\cal M}^{(1)}_8 = M_1 + M_2 = 
\langle D | \bar c (0) \gamma_\mu ( 1 + \gamma_5) t_a
\Bigl [ \langle \pi |
\bar d \gamma^\mu ( 1 + \gamma_5) u | 0 \rangle^a_8 \Bigr ]
b (0) | B \rangle = \nonumber \\
\frac{i g_s}{\pi^2} 
\int dz d^4 x
\frac{\phi(z)}{x^4} e^{izxq}
\Bigl [ x_\mu q_\nu - (x \cdot q) g_{\mu \nu} -
i \epsilon_{\mu \nu \alpha \beta} x_\alpha q_\beta \Bigr ]
\langle D | \bar c (0) \Gamma_\mu A_\nu (x) 
b (0) | B \rangle - \nonumber \\
\frac{i g_s}{\pi^2} 
\int dz d^4 x
\frac{\phi(z)}{x^4} e^{i(1-z)xq}
\Bigl [ x_\mu q_\nu - (x \cdot q) g_{\mu \nu} +
i \epsilon_{\mu \nu \alpha \beta} x_\alpha q_\beta \Bigr ]
\langle D | \bar c (0) \Gamma_\mu A_\nu (x) 
b (0) | B \rangle 
\end{eqnarray}
where $\bar \Gamma = \gamma_\mu (1 + \gamma_5)$ and
$t^a A^a(x) = A(x)$\footnote{Note that the transferred momentum $q$
can be written as $q = En$ where $n$
is a null vector (as in the approximation of collinear u and d - quarks) 
and $E$ is energy transferred to the
light quark system (it is large in the kinematical limit
under consideration). Observe that contrary to \cite{dg} 
there are terms in (\ref{octet}) which are
not proportional to $n \cdot A$.}.

This amplitude is also nonlocal, so we use the same idea as
in Eq. (\ref{expansion})
expanding it around $x^2 \sim 0$ and introducing ``generating 
functions''. It is clear that the only 
matrix element to be parameterized is 
\begin{equation}
\langle D | \bar c (0) \bar \Gamma_\mu A_\nu (x) 
b (0) | B \rangle 
\end{equation}
In FS gauge one can expand $A_\mu(x)$ about $x = 0$ using
``gauge invariant'' decomposition \cite{fs}
\begin{equation}
A_\mu (x) = \sum_{n=0}^{\infty} \frac{1}{(n+2)n!}
x^{\nu_1}...x^{\nu_n} x^{\rho}
\Bigl(D_{\nu_1}...D_{\nu_n} G_{\rho \mu} (0) 
\Bigr)
\end{equation}
This implies the matrix elements
\begin{eqnarray}
& \left.
\langle D | \bar c (0) \bar \Gamma_\mu A_\nu (x) 
b (0) | B \rangle = \right. & 
\nonumber \\
& \left.
\frac{i}{2 g_s} \sum_{n=0}^{\infty} \frac{1}{(n+2)n!}
x^{\nu_1}...x^{\nu_n} x^{\rho}
\langle D | \bar c(0) \bar \Gamma_\mu
D_{\nu_1}...D_{\nu_n}  \Bigl [ D_{\rho},D_{\nu} \Bigr ]
b (0) | B  \rangle = \right. &
\nonumber \\
& \left.
\frac{i}{2 g_s} \sum_{n=0}^{\infty} \frac{1}{(n+2)n!}
x^{\nu_1}...x^{\nu_n} x^{\rho} M_{\mu \nu_1 ... \nu_n \rho \nu}
\right. &
\end{eqnarray}
In order to parameterize the matrix elements appearing in the 
expression above we note that there are a total of $n+3$ indices 
and that the whole expression is
anti-symmetric with respect to $\rho \leftrightarrow \nu$ interchange.
Also, there are only two vectors (or their combinations) available,
so we choose them to be $p_B$ and $q$. This leads to
\begin{eqnarray} \label{param}
& \left.
M_{\mu \nu_1 ... \nu_n \rho \nu} =
\sum^n_{i=0} a_i {p_B}_\mu {p_B}_{\nu_1} ...{p_B}_{\nu_i}
q_{\nu_{i+1}} ...q_{\nu_n} \Bigl ( {p_B}_\rho q_\nu -
{p_B}_\nu q_\rho \Bigr ) + \right. &
\nonumber \\
& \left.
\sum^n_{i=0} b_i q_\mu {p_B}_{\nu_1} ...{p_B}_{\nu_i}
q_{\nu_{i+1}} ...q_{\nu_n} \Bigl ( {p_B}_\rho q_\nu -
{p_B}_\nu q_\rho \Bigr ) \right. &
\end{eqnarray}
with $a_i$ and $b_i$ being unknown coefficients.
Performing the contraction and 
dropping terms proportional to $q^2$ and $x^2$
we have for the first matrix element in (\ref{octet})
\begin{eqnarray} \label{first}
& \left.
M_1=
-\frac{1}{2 \pi^2} \int d^4x dz \frac{\phi(z)}{x^4} e^{izxq}
\sum_{n=0}^\infty \sum_{i=0}^n \frac{1}{(n+2)n!} 
\Bigl \{
-2 a_i (p_B q) (p_B x) (xq) - 
\right. &
\nonumber \\
& \left.
b_i (xq)^2 (p_B q) + a_i m_B^2 (xq)^2 + 
b_i (xq)^2 (p_B q) \Bigr \} (p_B x)^i (xq)^{n-i} 
\right. &
\end{eqnarray}   
Note that in this limit 
the dependence of the matrix element on $b_i$ 
cancels out.
These sums can be
parameterized in terms of some distribution function $f (n, \nu)$. Using 
the binomial formula, we obtain
\begin{eqnarray}
& \left.
\sum_{i=0}^n a_i (p_B x)^i (xq)^{n-i} =
\int (\nu p_B x + xq )^n f(n, \nu) d \nu 
\right. &
\nonumber \\
& \left.
\int d \nu \nu^i f (n, \nu) = 
\frac{a_i}{C^i_n} 
\right. &
\end{eqnarray}
where $C^i_n$ is a binomial coefficient. This gives for the matrix element
\begin{eqnarray}
M_1=-\frac{1}{2 \pi^2} \int d^4 x dz d \nu e^{izxq}
\Bigl[
m_B^2 (xq)^2 - 2 (p_B q) (p_B x) (xq) 
\Bigr]
\sum_n \frac{(\nu p_B x + qx)^n}{(n+2)n!}  
f (n, \nu)
\end{eqnarray}
Performing resummation with respect to $n$ we arrive at
\begin{eqnarray}
& \left.
\sum_n \frac{(\nu p_B x + qx)^n}{(n+2)n!}  
f (n, \nu) = 
\int d \alpha \psi (\alpha, \nu) e^{i \alpha(\nu p_B x + xq)}
\right. &
\nonumber \\
& \left.
\int d\alpha \alpha^n \psi (\alpha, \nu) = 
\frac{(-i)^n}{n+2} f (n, \nu) 
\right. &
\end{eqnarray}
This results in
\begin{eqnarray} \label{firt}
M_1= -\frac{1}{2 \pi^2} \int dz d \nu d \alpha
\phi (z) \psi (\alpha, \nu)
\Bigl[m_B^2 q_\mu q_\nu - 2 (p_B q) {p_B}_\mu q_\nu \Bigr ]
\int d^4 x \frac{x_\mu x_\nu}{x^4} e^{i (\alpha \nu p_B  + 
(\alpha + z) q)x}
\end{eqnarray}
Similarly, for the second matrix element one obtains
\begin{eqnarray} \label{second}
M_2= \frac{1}{2 \pi^2} \int dz d \nu d \alpha
\phi (z) \psi (\alpha, \nu)
\Bigl[m_B^2 q_\mu q_\nu - 2 (p_B q) {p_B}_\mu q_\nu \Bigr ]
\int d^4 x \frac{x_\mu x_\nu}{x^4} e^{i (\alpha \nu p_B  + 
(1 + \alpha + z)q)x}
\end{eqnarray}
At this stage we have classified the infinite number of the 
unknown matrix elements in terms of a few continuous functions. The
strategy at this point in general would be to consider the structure
of these functions and make reasonable smooth guesses as to their form.
For example $\phi (z)$ is related to wavefunction used in the studies
of the perturbative limit of the pion form factor and has the
asymptotic form $\phi(z) = 2 f_\pi z (1-z)/\sqrt{6}$\cite{brodsky}. 
However, we find
that an important cancellation occurs which is independent of the form
of these structure functions.
Taking $a_\beta = (\alpha + z) q_\beta + \alpha \nu {p_B}_\beta$,
$b_\beta = (1+ z + \alpha) q_\beta + \alpha \nu {p_B}_\beta$
and  integrating over $x$ we obtain for the 
${\cal M}_8^{(1)}$:\footnote{The key integral is
$$ I_{\mu \nu} = - \frac{\partial}{\partial a_\mu}
\frac{\partial}{\partial a_\nu} 
\int d^4 x  \frac{e^{i ax}}{x^4} = 
2i\pi^2 \Bigl( -2 \frac{a_\mu a_\nu}{a^4} + \frac{g_{\mu \nu}}{a^2}
\Bigr)
$$}

\begin{eqnarray}
& \left.
{\cal M}^{(1)}_8 = 
-i \int d \nu d \alpha \psi_1 (\alpha, \nu)  
\int d z \phi(z) \right. &
\nonumber \\
& \left.
\Bigl \{
\frac{1}{a^4}
\Bigr[ 4 (p_B q) (p_B a) (q a) -
2 m_B^2 (qa)^2 - 2 a^2 (p_B q)^2 \Bigr]- 
\right. &
\nonumber \\
& \left.
\frac{1}{b^4}
\Bigr[4(p_B q) (p_B b) (q b) -
2 m_B^2 (qb)^2 - 2 b^2 (p_B q)^2 \Bigr] 
\Bigr \}
= 0
\right. &
\end{eqnarray}
at the leading order in $q^2$ for any $\phi$ and $\psi$. This 
implies that the semi-soft-one-gluon matrix elements do not bring
new contributions to the factorization approximation result in
the kinematical limit under consideration. 

We would like to point out that the two-pseudoscalar
final state is somewhat special: for instance, $B \to D \rho$ does not
fall into the same category - there is a good chance that 
under the similar conditions there is no complete cancellation of the 
diagrams. This is due to the fact that there can be extra terms in 
(\ref{param}) that include polarization vectors of $\rho$ meson.

In conclusion,
we have constructed a QCD-based model for the 
factorization approximation in the limit 
$E \sim {m_b}_{m_b \to \infty}$. In the framework of the model 
the complete cancellation of the non-factorizable terms is
observed at the leading order in $q^2,~x^2$. The OPE is resummed 
by introducing two generating (distribution) functions that
can be independently parameterized. The model gives a consistent
way to estimate non-leading (non-factorizable) contributions.

\begin{figure}[t]
\centering
\leavevmode
\centerline{
\epsfbox{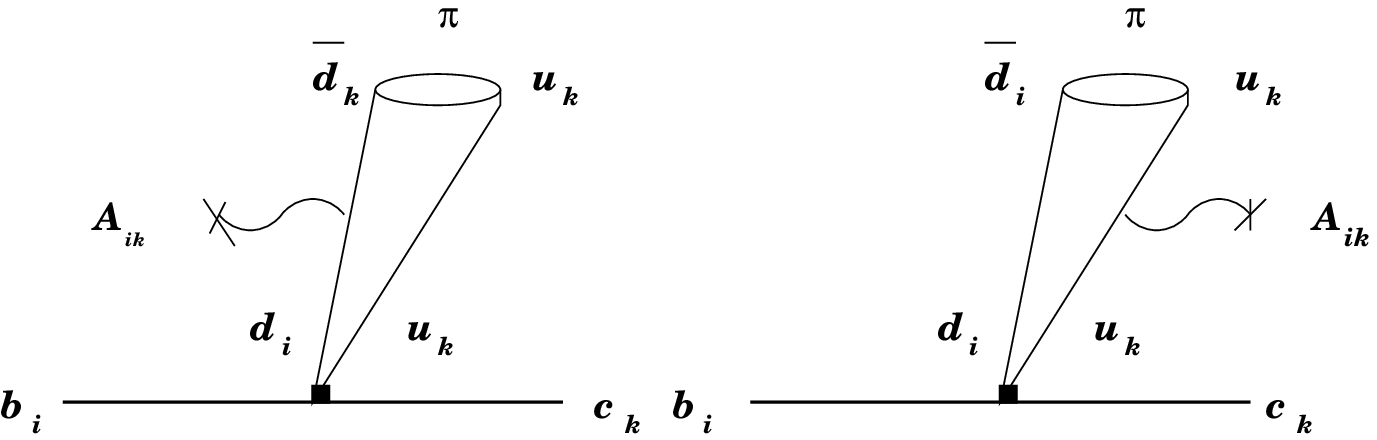}}
\caption{One gluon diagrams for the $B \to D \pi$.}
\end{figure}

\end{document}